\documentclass{aastex}
\usepackage{emulateapj5}
\usepackage{graphicx}
\usepackage{times,psfig}
\usepackage{natbib}

 \usepackage{url}
\def\spose#1{\hbox to 0pt{#1\hss}}
\newcommand\lsim{\mathrel{\spose{\lower 3pt\hbox{$\mathchar"218$}}
     \raise 2.0pt\hbox{$\mathchar"13C$}}}
\newcommand\gsim{\mathrel{\spose{\lower 3pt\hbox{$\mathchar"218$}}
     \raise 2.0pt\hbox{$\mathchar"13E$}}}
\def\ltsima{$\; \buildrel < \over \sim \;$}
\def\lsim{\lower.5ex\hbox{\ltsima}}
\def\gtsima{$\; \buildrel > \over \sim \;$}
\def\gsim{\lower.5ex\hbox{\gtsima}}

\def\apjl{ApJL}
\def\apjs{ApJSS}

\def\aap{A\&A}
\def\sch{Schwarzschild}
\def\ergcms{{\rm\thinspace erg \thinspace cm^{-2} \thinspace s^{-1}}}
\def\ergs{{\rm\thinspace erg \thinspace s^{-1}}}

\shorttitle{B2~1023+25 observed by {\it NuSTAR}}
\shortauthors{Sbarrato et al.}

\begin{document}
\bibliographystyle{plainnat}

\title{{\it NuSTAR} detection of the blazar B2~1023+25 at redshift 5.3 
	}
\author{T.\  Sbarrato\altaffilmark{1,2},  
G.\ Tagliaferri\altaffilmark{2}, G.\ Ghisellini\altaffilmark{2}, 
M.\ Perri\altaffilmark{3,4}, S.\ Puccetti\altaffilmark{3,4}, 
M.\ Balokovi\'{c}\altaffilmark{5}, M.\ Nardini\altaffilmark{6}, 
D.\ Stern\altaffilmark{7}, 
S.\ E.\ Boggs\altaffilmark{8}, W.\ N.\ Brandt\altaffilmark{9,10}, 
F.\ E.\ Christensen\altaffilmark{11}, P.\ Giommi\altaffilmark{3,4}, 
J.\ Greiner\altaffilmark{12}, C.\ J.\ Hailey\altaffilmark{13}, 
F.\ A.\ Harrison\altaffilmark{5}, T.\ Hovatta\altaffilmark{5}, 
G.\ M.\ Madejski\altaffilmark{14}, 
A.\ Rau\altaffilmark{12}, P.\ Schady\altaffilmark{12}, V.\ Sudilovsky\altaffilmark{12}, 
C.\ M.\ Urry\altaffilmark{15}, 
W.\ W.\ Zhang\altaffilmark{16}
}
\altaffiltext{1}{Dipartimento di Scienza e Alta Tecnologia, 
			Universit\`a dell'Insubria, Via Valleggio 11, I--22100 Como, Italy; 
			tullia.sbarrato@brera.inaf.it}
\altaffiltext{2}{INAF -- Osservatorio Astronomico di Brera, via E. Bianchi 46, I--23807 Merate, Italy.}
\altaffiltext{3}{ASI -- Science Data Center, via Galileo Galilei, I--00044 Frascati, Italy}
\altaffiltext{4}{INAF -- Osservatorio Astronomico di Roma, via Frascati 33, 
		      	I--00040 Monteporzio Catone, Italy}
\altaffiltext{5}{Cahill Center for Astronomy and Astrophysics, California Institute
			of Technology, Pasadena, CA 91125, USA}
\altaffiltext{6}{Dipartimento di Fisica G.\ Occhialini, Universit\`a di Milano Bicocca, 
			Piazza della Scienza 3, I--20126 Milano, Italy}
\altaffiltext{7}{Jet Propulsion Laboratory, California Institute of Technology, Pasadena, 
			CA 91109, USA}
\altaffiltext{8}{Space Sciences Laboratory, University of California, Berkeley, CA 94720, USA}
\altaffiltext{9}{Department of Astronomy \& Astrophysics, The Pennsylvania State University, 
			525 Davey Lab, University Park, PA 16802, USA}
\altaffiltext{10}{Institute for Gravitation and the Cosmos, The Pennsylvania State University, 
			University Park, PA 16802, USA}
\altaffiltext{11}{DTU Space - National Space Institute, Technical University of
			Denmark, Elektrovej 327, 2800 Lyngby, Denmark}
\altaffiltext{12}{Max--Planck--Institut f\"ur extraterrestrische Physik, Giessenbachstrasse 1, 
			85748, Garching, Germany}
\altaffiltext{13}{Columbia Astrophysics Laboratory, Columbia University, New York, 
			NY 10027, USA} 
\altaffiltext{14}{Kavli Institute for Particle Astrophysics and Cosmology,
			SLAC National Accelerator Laboratory, Menlo Park, CA 94025, USA}
\altaffiltext{15}{1Department of Physics, Yale University, New Haven, CT 06520-8121, USA}
\altaffiltext{16}{NASA Goddard Space Flight Center, Greenbelt, MD 20771, USA}


\received{~~} \accepted{~~}
\journalid{}{}
\articleid{}{}
\authoremail{tullia.sbarrato@brera.inaf.it}

\begin{abstract}

B2~1023+25 is an extremely radio--loud quasar at $z = 5.3$  
which was first identified as a likely high--redshift blazar candidate 
in the SDSS+FIRST quasar catalog.
Here we use the {\it Nuclear Spectroscopic Telescope Array} ({\it NuSTAR}) 
to investigate its non--thermal jet emission, whose high--energy component we detected
in the hard X--ray energy band.
The X--ray flux is $\sim 5.5 \times 10^{-14}\ergcms$ 
(5-10~keV) and the photon spectral index is $\Gamma_{\rm X} \simeq 1.3-1.6$.
Modeling the full spectral energy distribution, we find that the jet is oriented 
close to the line of sight, with a viewing angle of $\sim 3^{\circ}$, and has significant
Doppler boosting, with a large bulk Lorentz factor $\sim 13$, 
which confirms the identification of B2~1023+25 as a blazar.
B2~1023+25 is the first object at redshift larger than 5 detected by {\it NuSTAR}, 
demonstrating the ability of {\it NuSTAR} to investigate the early X--ray Universe and
to study extremely active supermassive black holes located at very high redshift.

\end{abstract}

\keywords{galaxies: active -- quasars: general -- X--rays: general --- 
		quasars: individual: B2~1023+25}


\section{Introduction}
\label{sec-intro}

Blazars are radio--loud Active Galactic Nuclei (AGN) with a 
relativistic jet directed at a small angle from our line of sight 
(Urry \& Padovani 1995).
The peculiar orientation relativistically boosts  
the radiation emitted from their jets, 
making them visible even at high redshift ($z$).

The typical Spectral Energy Distribution (SED) of a blazar is 
dominated by its non--thermal emission, characterized 
by two broad humps: the lower frequency component is attributed 
to synchrotron emission, while the higher frequency component 
is attributed to Inverse Compton (IC) emission.
The humps of a blazar SED peak at lower frequencies as the bolometric 
luminosity increases, at least for blazars studied to date 
(see e.g.\ Fossati et al.\ 1998; Ghisellini et al.\ 2011 and Giommi et al.\ 2012 for different 
interpretations of the effect).
At $z>4$ we expect to see only the most powerful objects.
Therefore we should detect high--redshift sources whose synchrotron
hump peaks in the sub--mm, and 
whose IC hump peaks in the $\sim$MeV band.
The synchrotron component shifts far enough out of the optical/UV to leave the accretion 
disk, usually swamped by non-thermal emission, visible (Sbarrato et al.\ 2013; Wu et al.\ 2013).
The optical emission of a high--power, high--redshift blazar therefore becomes 
indistinguishable from that of a high--redshift,  radio--quiet quasar.
Since the accretion disk is visible, an estimate of the 
black hole mass can be obtained by fitting the emitted spectrum with a 
Shakura \& Sunyaev (1973) accretion disk model (Calderone et al.\ 2013; 
Sbarrato et al.\ 2013).

The large redshift ($z>4$)  also moves the observed SED to lower frequencies.
Therefore, the peak of the high--energy component appears well below 100 MeV 
(Ghisellini et al.\ 2010a; 2010b; Sbarrato et al.\ 2012).
This introduces a challenge in identifying and classifying
high--power, high--redshift blazars, since a classic hallmark of a blazar is
detection by a $\gamma$--ray instrument
like the Large Area Telescope (LAT) on the {\it Fermi} 
satellite (Atwood et al.\ 2009).  
Since the high--energy component peaks below 100 MeV, 
the identification of powerful high--redshift blazars in the $\gamma$-ray is problematic.

On the other hand, sensitive hard X--ray telescopes can detect the 
high--energy hump of high--redshift, extremely powerful blazars.
In fact, the Burst Alert Telescope (BAT) onboard the {\it Swift} satellite 
(Gehrels et al.\ 2004) has detected blazars up to larger redshift
than the LAT (Ajello et al.\ 2009).
The BAT blazars have hard X--ray spectra
[i.e. $\alpha_X \lsim 0.5$, assuming $F(\nu) \propto \nu^{-\alpha_{x}}$], 
and this, together with a strong X--ray to optical flux ratio, can be generally 
taken as a signature of the blazar nature of a source.

The most distant blazar known is Q0906+6930 (Romani et al.\ 2004; Romani 2006), 
located at $z=5.47$. 
It was first classified as a blazar through a serendipitous EGRET $3\sigma$ detection.
However the more sensitive {\it Fermi}/LAT instrument did not detect the source, 
so the EGRET detection
could either be spurious, or due to an episode of exceptional activity.
Subsequent to its classification, Q0906+6930 was confirmed as a blazar 
through its X--ray and radio features (Romani 2006).
In addition to being radio-loud, this source indeed shows a hard X--ray spectral index and high X-ray flux,
leading to its classification as the most distant known blazar.

We identified the second most distant blazar 
known, 
B2~1023+25 at $z=5.3$ (Sbarrato et al.\ 2012, hereafter S12).
Its extremely large mass, $M_{\rm BH}=2.8\times10^9M_\odot$, 
derived by fitting the accretion disk spectrum (see S12),
makes this object particularly interesting, since it is possible to put relevant 
constraints on supermassive black hole formation models using B2~1023+25 
as a tracer of a population of very high--redshift, extremely massive black holes. 
Indeed, 
the observation of a single blazar with viewing angle $\theta_{\rm v}$ smaller than or comparable to the 
jet beaming angle (i.e.\ $\theta_{\rm v}<1/\Gamma$, where $\Gamma$ is the bulk 
Lorentz factor of the relativistic jet) indicates the possible presence of $2\Gamma^2$ 
analogous radio--loud, extremely massive AGN with their jets directed in random directions. 
A typical blazar has $\Gamma\sim15$, so finding even a few blazars at very high 
redshift is statistically very important for studying the population of extremely 
massive black holes. Therefore this line of research could become competitive with searches for
supermassive black holes at high redshifts using radio--quiet quasar samples: 
each blazar with $M_{\rm BH}>10^9M_\odot$ implies the presence of 
hundreds of analogous black holes in systems with a jet pointing elsewhere. 
Note that the usual radio--loud to radio--quiet ratio (10\%) refers to objects 
with any black hole mass. 
At the high--mass end, and at high redshift, this ratio could be larger 
(Volonteri et al.\ 2011; Ghisellini et al.\ 2013).
The existence of $z>5$ massive ($M_{\rm BH}>10^9M_\odot$) black holes in sources with
powerful jets also raises challenges for BH growth models.   Rapidly-spinning (Kerr) BHs
are often invoked as the energy source behind powerful jets
(e.g.\ Wilson \& Colbert 1995; Sikora, Stawarz \& Lasota 2007). 
However, their accretion disks are radiatively very efficient 
($L_{\rm d}=\eta\dot Mc^2$, with $\eta>0.1$, up to 0.3; Thorne 1974).
As a consequence, over the same time interval, 
Eddington--limited Kerr black holes accrete {\it less matter} 
than Schwarzschild BHs.  If this is the case, 
the time needed to form a $M_{\rm BH}\ge10^9M_\odot$ black hole
in a source with powerful jets is very long, and no such systems should exist at 
$z>4$ if they grow primarily through persistent accretion (Ghisellini et al.\ 2013). 

B2~1023+25 was selected from a sample of $z>4$ radio--loud sources 
as the best blazar candidate (Sbarrato et al.\ 2013; 
note that it was also listed as one of the most radio--loud quasars at $z>4$ by Wu et al.\ 2013).
We were able to observe the rising part of its high--energy hump thanks 
to a ToO observation with the X--Ray Telescope (XRT; Burrows et al.\ 2005) 
onboard the {\it Swift} satellite (Gehrels et al.\ 2004).
The hardness of the X--ray spectrum suggested that B2~1023+25 is a bona fide blazar.
However {\it Swift}/XRT observes at frequencies too low 
to properly sample the high--energy hump.
This motivated our observation of B2~1023+25 with the 
{\it Nuclear Spectroscopic Telescope Array} ({\it NuSTAR}, Harrison et al.\ 2013).
Indeed, broader X-ray bandpass is required to properly classify B2~1023+25 by
observing closer to the peak of the inverse Compton component.
Thanks to its unparalleled broad--band sensitivity,  {\it NuSTAR} is an ideal instrument to determine if 
the X--ray spectrum and flux of high--redshift candidates  
are respectively hard and intense enough to 
classify them as powerful blazars.

Here we present {\it NuSTAR} observations of B2~1023+25 along with  
simultaneous observations obtained in multiple energy bands: 
X--ray observations from  {\it Swift}/XRT, 
radio observations at three different frequencies from CARMA and OVRO, 
and seven--band optical--NIR photometry from La Silla Observatory in Chile 
with GROND (\S  \ref{sec-data}).
The X--ray data in particular allows us to constrain the high--energy component, 
thereby providing important insights into the orientation 
and Compton--boosting of B2~1023+25 (\S \ref{sec-discussion}). 

In this work, we adopt a flat cosmology with $H_0=70$ km s$^{-1}$ Mpc$^{-1}$ and
$\Omega_{\rm M}=0.3$.

\section{Observations and Data Analysis}
\label{sec-data}

We performed simultaneous observations of B2~1023+25 with five instruments 
at different frequencies on UT 2013 January 1.
The X--ray band was covered with {\it Swift}/XRT to check for possible variability 
with respect to our previous observations, and {\it NuSTAR} to study the 
hard X--ray energy range.
We added data from a previous {\it Chandra} observation (Wu et al. 2013)
in order to increase the statistics of the soft X--ray energy band.
We re--observed the source from La Silla (Chile) with the Gamma--Ray Burst 
Optical Near--Infrared Detector (GROND; Greiner et al.\ 2008), which provides 
simultaneous photometric data from the NIR to the optical in 7 different bands.
This re--observation was performed to check for possible variability 
of the thermal emission of the source.
To have simultaneous radio data at three different frequencies, we observed 
B2~1023+25 with the Combined Array for Research in Millimeter--wave Astronomy 
(CARMA; Bock et al.\ 2006) and with the 40--meter telescope at the 
Owens Valley Radio Observatory (OVRO). 
CARMA observed at 31 and 91 GHz (1 cm and 3 mm), 
while OVRO provided data at 15 GHz (2 cm).
These data are combined with data from the {\it Wide--field Infrared Survey Explorer} 
({\it WISE}\footnote{Data retrieved from the {\it WISE} All--Sky Source Catalog: 
\url{http://irsa.ipac.caltech.edu/.}}) 
satellite (Wright et al.\ 2010) and with 
archival data from NASA/IPAC Extragalactic 
Database (NED) and the ASI Science Data Center 
(ASDC\footnote{\url{http://tools.asdc.asi.it/.}}).

\subsection{X--ray observations}

\subsubsection{{\it NuSTAR} observations}

The {\it NuSTAR} satellite observed B2~1023+25 beginning on UT 2012 December 31 
(sequence 60001107002) for a net exposure time of 59.3 ks. 
The two data sets obtained with the {\it NuSTAR} Focal Plane 
Modules A and B (FPMA and FPMB) 
were first processed with the NuSTARDAS software
package (v.1.2.0) jointly developed by the ASI Science Data Center 
(ASDC) and the California Institute of Technology (Caltech). 
Event files were calibrated and cleaned with standard filtering
criteria with the {\tt nupipeline} task using version 20130509 of
the {\it NuSTAR} Calibration Database (CALDB).

The FPMA and FPMB spectra were extracted from the
cleaned event files using a circular aperture of 12 pixel ($\sim30$'') radius,
while the background was extracted from two distinct nearby circular
regions of 30 pixel radius. The ancillary response files were
generated with the {\tt numkarf} task, applying corrections for the PSF
losses, exposure maps and vignetting. 
The source was detected up to 20 keV, and 
the source spectrum in the $4-20$
keV energy band was formed from a total of 79 counts (of which $\sim44$ are from the
background) for FPMA and 113 counts (of which $\sim58$ are from the background) for
FPMB. Both spectra were binned to ensure a minimum of 1 count per
bin.

\subsubsection{{\it Swift} observations}

The {\it Swift} satellite observed the source three times: on UT 2012 June
21 (sequence 00032500001), on UT 2012 June 22 (sequence 00032500002) and
on UT 2012 December 31 (sequence 00080499001). 
All XRT observations were
carried out using the most sensitive Photon Counting (PC) readout
mode.

The XRT data set was first processed with the XRTDAS software package
(v.2.8.0) developed at the ASDC and
distributed by HEASARC within the HEASoft package (v. 6.13). Event
files were calibrated and cleaned with standard filtering criteria
with the {\tt xrtpipeline} task using the latest calibration files
available in the {\it Swift} CALDB.

The spectra obtained from the single observations 
are perfectly consistent, with an uncertainty on each measurement of $\sim20-25\%$, 
showing no variability among the three observations.   
We therefore merge the individual XRT event files, using the 
XSELECT package for a total net exposure time of 20.3 ks.  Next we extracted the
average spectrum from the summed cleaned event
file. Events for the spectral analysis were selected within a circular aperture 
of 10 pixel ($\sim23$'') radius, which encloses about 80\% of the PSF,
centered on the source position. The background was extracted from a
nearby circular region of 100 pixel radius. The ancillary response
files were generated with the {\tt xrtmkarf} task applying corrections
for the PSF losses and CCD defects using the cumulative exposure
map. The latest response matrices available in the {\it Swift} CALDB were
used. The source spectrum in the $0.3-10$ keV energy band was formed from 
a total of 41 counts (of which $\sim3$ are from the background) and it was binned to
ensure a minimum of 1 count per bin.

\subsubsection{{\it Chandra} observations}

B2~1023+25 was observed by {\it Chandra} on 2011 March for a total
of $\sim$5 ksec with the ACIS camera. 
These data were presented in Wu et al. (2013). In order to use
them together with our other data sets, we re-extracted the {\it Chandra} spectrum.
The data were reduced with the CIAO
4.4 package (Fruscione et al. 2006) using the {\it Chandra} 
CALDB version 4.4.7, adopting standard procedures. The
source spectrum was extracted in a circular region centered on the
peak of the X-ray source emission and with a radius of
3". The background spectrum was extracted from four circular regions
with $\sim$5" radii, located around the source. The source spectrum
in the $0.5-7$ keV energy band was formed by a total of 54 counts (of which $\sim1$
is from the background) and it was binned to ensure a minimum of 1 count
per bin.

\subsubsection{X--ray spectral analysis}
\label{xray_an}

A comparison of the current and previous {\it Swift}/XRT observations discussed in S12
show that the source did not vary between these epochs. 
If we fit the data of each satellite alone, we find a good agreement among 
{\it NuSTAR}, XRT and {\it Chandra}, but due to the faintness of the source the
uncertainties are quite large. 
Since there is no evidence for variability we performed a
simultaneous fit of the {\it Swift}/XRT, {\it Chandra} and {\it NuSTAR} spectra 
using the XSPEC package, and adopting C--statistics (Cash 1979). 

In prior work the XRT and {\it Chandra} data were fit with a simple power law model plus
Galactic absorption (S12, Wu et al. 2013). 
In the present analysis, if we leave  $N_{\rm H}$ free
to vary, we find a high value for the absorption and a slightly steeper spectrum than previously
reported.  The statistical quality of the data is, however, not sufficient to distinguish
between no absorption and spectral curvature and a higher level of intervening absorption
and steeper spectral index.  X--ray absorption in the host of blazars is unlikely 
(the jet is able to completely
ionize  host ISM), but intervening material in high--redshift objects 
could be responsible for extra absorption. 
The presence or absence of extra absorption due to intervening material 
in quasar X--ray spectra is a matter of debate 
(see e.g.\ Vignali et al.\ 2005; Shemmer et al.\ 2005, 2006; 
Yuan et al.\ 2006; for a different point of view, see Behar et al.\ 2011), so we investigate fits with
the column both fixed to the Galactic value and free to vary.
Both models provided a good description of the observed spectra (see Table \ref{Xspec}).
If we leave $N_{\rm H}$ free to vary, we find a high value of 
$2.8 \times 10^{21}$ cm$^{-2}$ assuming
the absorbing material is at $z=0$ (for higher redshift absorbers this column increases).
A simple power--law plus Galactic absorption
also provides an acceptable fit to the data: $\chi ^2$=230.5 for 253 degree of freedom, to be compared
with $\chi ^2$=211.3 for 252 degree of freedom for the case with $N_{\rm H}$ free to vary.
Although the $\chi ^2$ clearly improves when $N_{\rm H}$ is free to vary,  the reduced $\chi ^2$
is already $<1$ with $N_{\rm H}$ fixed to the Galactic value;  therefore, we cannot discriminate
between the two possibilities from a statistical point of view.
The results of the spectral fits are shown in Table \ref{Xspec}.  
The photon spectral index in the two cases varies from 
$\Gamma_{\rm X} = 1.29^{+0.14}_{-0.15}$ to $\Gamma_{\rm X} = 1.60^{+0.27} _{-0.26}$.
In the following  analysis we take this level of uncertainty in the spectral index into account, 
in particular
in constraining the jet viewing angle (see \S \ref{sec-discussion}).
In Figure \ref{zoom} we plot the X-ray SED of the source as derived
with the spectral fit performed with $N_{\rm H}$ free to vary.

\begin{table*} 
\centering
\begin{tabular}{lllllll}
\hline
\hline
 $N_{\rm H}$ & $F_{\rm norm}$ & $\Gamma_{\rm X}$ & $F_{\rm 5-10kev}$  & $\chi^2$ / dof \\
(${\rm cm}^{-2}$) & (${\rm ph\,cm^{-2}\,s^{-1}}$) & & ($\ergcms$) & \\
\hline   
   $1.5 \times 10^{20}$ fixed & $1.29_{-0.26}^{+0.29}\times10^{-5}$ 
				& $1.29^{+0.14}_{-0.15}$ & $5.8\times10^{-14}$ & 230.5 / 253\\
 $2.8_{-1.7}^{+2.0}\times 10^{21}$ & $2.26_{-0.77}^{+1.16}\times10^{-5}$ 
				& $1.60^{+0.27}_{-0.26}$ & $5.5\times10^{-14}$ & 211.3 / 252 \\
\hline
\hline 
\end{tabular}
\caption{Parameters of the X--ray spectral analysis. The errors are at 90\% level of confidence for one parameter of interest.
}
\label{Xspec}
\end{table*}

\subsection{GROND observations}
\label{sec-grond}

The 7--band  GROND imager, mounted at the 2.2 m MPG/ESO telescope at La Silla Observatory
(Chile), started observing B2~1023+25 on UT 2013 January 1 at 07:38:43 UTC. 
We carried out three 8--minute observations simultaneously in all 
seven $g^\prime, r^\prime, i^\prime, z^\prime, J, H, K_{\rm s}$ bands for a total exposure 
time of 1379 s in the optical and 1440 s in the near--IR (NIR) bands. 
Observations were carried out at an average seeing of 1.2'' evaluated 
from the $r^\prime$--band image, and at an average airmass of 1.8. 
The source was  clearly detected in all bands but $g^\prime$ for which an upper limit of 
23.4 (AB magnitude) was found.

The GROND  optical and NIR image reduction and photometry were
performed using standard IRAF tasks (Tody 1993), similar to the
procedure described  in Kr\"uhler et al. (2008). 
A general model for the point--spread function (PSF) of each image was
constructed using bright field stars, and it was then fitted to the point source. 
The absolute calibration of the $g^\prime, r^\prime, i^\prime, z^\prime$ bands was 
obtained with respect to the magnitudes of SDSS stars within the 
blazar field while the $J, H, K_{\rm s}$ 
bands calibration was obtained with respect to magnitudes of 
Two Micron All Sky Survey (2MASS) stars (Skrutskie et al. 2006).

Table \ref{grond} reports the observed AB magnitudes, not corrected for 
the Galactic extinction of $E(B-V)=0.02$ from Schlegel et al. (1998).
Note that these data are fully consistent with those from S12, showing that the 
thermal emission of the source  did not vary
between the two observing time, 
as was also found for the X--ray 
non--thermal emission (see \S \ref{xray_an}). 

\begin{table*} 
\centering
\begin{tabular}{llllllll}
\hline
\hline
~  &$g^\prime$ &$r^\prime$ &$i^\prime$ &$z^\prime$ &$J$ &$H$ &$K_{s}$ \\
\hline   
$\lambda_{\rm eff}$ (\AA)   &4587 &6220 &7641 &8999 &12399 &16468 &21706 \\  
AB magnitude      &$>$23.4 &22.16$\pm$0.16 &19.91$\pm$0.04 &19.73$\pm$0.03 &19.52$\pm$0.06 &19.20$\pm$0.09 &19.35$\pm$0.24 \\  
\hline
\hline 
\end{tabular}
\vskip 0.4 true cm
\caption{GROND AB observed magnitudes of B2~1023+25, taken UT 2013 January 1  
(magnitudes not corrected for Galactic extinction). 
The first row gives the effective wavelength of the filter (in Angstroms).
}
\label{grond}
\end{table*}

\subsection{CARMA observations}

We observed B2~1023+25 at 31 and 91~GHz (1~cm and 3~mm bands,
respectively) with CARMA. The observations were carried out
simultaneously with the {\it NuSTAR} observation on UT 2013 January 1,
with the array in the SL conÞguration. This conÞguration includes
eight 3.5-meter antennae on baselines of 5 to 85 meters. Single
sideband receivers were used to observe the upper and the lower
sideband at 3~mm and 1~cm, respectively, and the correlator was
configured to process 8~GHz of bandwidth. After flagging the bad data
intervals, the total observation time was 3~hours in each band. Strong
nearby sources 0956+252, 0927+390 and planet Jupiter were used as
gain, passband and ßux calibrators. The data were processed using the
Multichannel Image Reconstruction Image Analysis and Display (MIRIAD;
Sault, Teuben \& Wright 1995) software, optimized for CARMA. The
observations reached an RMS of 0.7~(1.5)~mJy in the 1~cm (3~mm) band,
providing a 47$\sigma$ (9$\sigma$) detection of the target. The
absolute ßux density calibration, however, adds a systematic
uncertainty of 10\%, so the ßux density values used in further
analysis are $f_{\nu}(31 {\rm GHz})=33\pm4$~mJy and
$f_{\nu}(91{\rm GHz})=14\pm3$~mJy.

\subsection{OVRO 40--meter observations}

The OVRO 40--meter telescope obtained a 15 GHz observation of B2~1023+25 
simultaneous with {\it NuSTAR} on UT 2013 January 1.  
The telescope uses off-axis dual-beam optics and a cryogenic high electron 
mobility transistor (HEMT) low-noise amplifier with a 3~GHz bandwidth. 
The two sky beams are Dicke switched using the off-source beam as a reference, 
and the source is alternated between the two beams in an ON-ON fashion 
to remove atmospheric and ground contamination. 
A noise level of approximately 3--4~mJy in quadrature with about 2\% 
additional uncertainty mostly due to pointing errors, 
is achieved in a 70-second integration cycle. 
The weighted average of 9 consecutive integrations was used to derive the 15~GHz 
flux density $f_{\nu}(15\mbox{GHz})=55\pm4$~mJy, where the systematic uncertainty 
in the absolute flux calibration has already been included. 
Calibration is routinely achieved using a temperature-stable diode noise source 
to remove receiver gain drifts and the flux density scale was derived from observations 
of 3C~286 assuming the Baars et~al.~(1977) value of 3.44~Jy at 15~GHz. 
Details of the reduction and calibration procedure can be found in Richards et~al.~(2011).

\begin{figure*}
\vskip -1 true cm 
\epsscale{1.}
\plotone{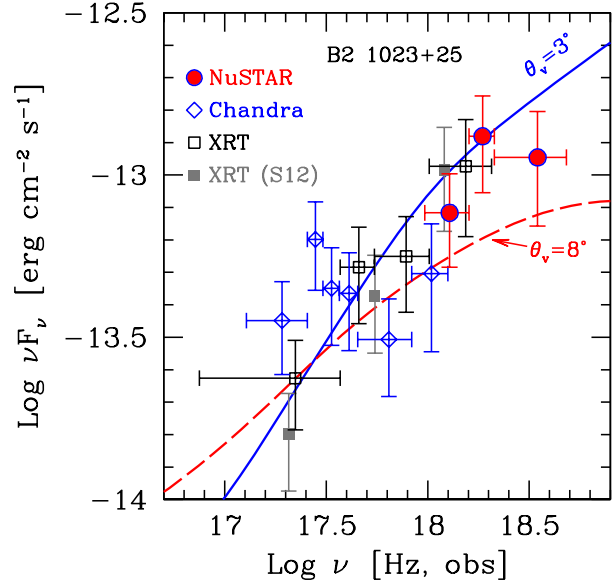}
\vskip -0.5 true cm
\caption{
X--ray spectrum of B2~1023+25, along 
with the two SED models discussed in the text.
{\it NuSTAR}/FPMB data are filled circles (red, circled in blue in the electronic version);  
{\it Chandra} data are empty diamonds, while {\it Swift}/XRT data are empty squares 
(respectively blue empty diamonds and black empty squares in the electronic version).
The solid (blue) line is the model with $\theta_{\rm v}=3^\circ$, $\Gamma=13$ and 
 parameters as in the first row of Table \ref{para}. 
The dashed (red) line is the model with $\theta_{\rm v}=8^\circ$, $\Gamma=10$ and 
parameters as in the second row of Table \ref{para}. 
} 
\label{zoom}
\end{figure*}

\begin{figure*}
\vskip -1 true cm 
\epsscale{1.4}
\plotone{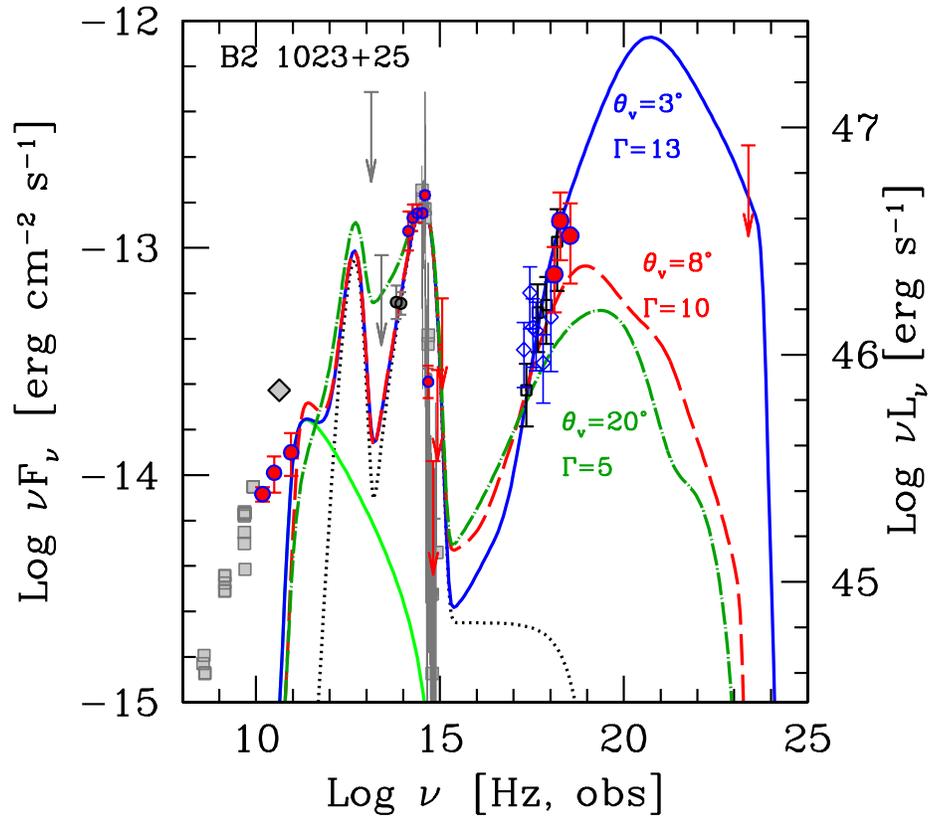}
\vskip -0.5 true cm
\caption{
Broad--band SED of B2~1023+25 together with the models discussed in the text.
Simultaneous OVRO, CARMA, GROND and {\it NuSTAR} data are filled circles 
(red points circled in blue in the electronic version).
{\it Chandra} data are empty diamonds, while {\it Swift}/XRT data are empty squares 
(respectively blue empty diamonds and black empty squares in the electronic version).
The (grey) filled symbols are data from literature: 
squares are archival data from ASDC,  
the diamond is the radio point from Frey et al.\ 2013, 
the circles and the two upper limits are {\it WISE} data, 
the line is the SDSS spectrum. 
The dotted (black) line is the thermal emission of the source, including the 
accretion disk, torus and X--ray corona emission.
The {\it Fermi}/LAT upper limit is for 3.8 years, 5$\sigma$ (red arrow).
The solid (blue) line is the model with parameters as in the first row of Table \ref{para}. 
The dashed (red) line is the model with parameters as in the second row of Table \ref{para}.
The dot--dashed (green) line is the model with parameters as 
in the third row of Table \ref{para}.
} 
\label{sed1}
\end{figure*}

\begin{table*}
\centering
\begin{tabular}{lllll lllll lllll }
\hline
\hline
$\Gamma$  &$\theta_{\rm v}$ &$R_{\rm diss}$ &$R_{\rm diss}/R_{\rm S}$   &$P^\prime_{\rm i}$ &$B$
&$\gamma_{\rm b}$ &$\gamma_{\rm max}$ &$s_1$  &$s_2$
&$\log P_{\rm r}$ &$\log P_{\rm B}$ &$\log P_{\rm e}$ &$\log P_{\rm p}$ \\
~[1]       &[2] &[3] &[4] &[5] &[6] &[7] &[8] &[9] &[10] &[11] &[12] &[13] &[14] \\
\hline
13  &3   &504  &600  &0.01  &2.3  &70 &4e3 &0    &2.6 &45.70 &45.93 &44.16 &46.61  \\
10  &8   &420  &500  &0.23  &4.4  &20 &4e3 &--1  &2.6 &46.72 &46.11 &45.72 &48.26  \\
5   &20  &588  &700  &7.0   &5.5  &2e3 &4e3 &--1  &2.6 &47.42 &45.98 &45.28 &47.73 \\
\hline
\hline
\end{tabular}
\vskip 0.4 true cm
\caption{Input parameters used to model the SED.
Col. [1]: bulk Lorentz factor;
Col. [2]: viewing angle (degrees);
Col. [3]: distance of the blob from the black hole in units of $10^{15}$ cm;
Col. [4]: $R_{\rm diss}$ in units of the \sch\ radius;
Col. [5]: power injected in the blob calculated in the comoving frame, in units of $10^{45}$ erg s$^{-1}$;
Col. [6]: magnetic field in Gauss;
Col. [7], [8] and: minimum and maximum random Lorentz factors of the injected electrons;
Col. [9] and [10]: slopes of the injected electron distribution [$Q(\gamma)$] below
and above $\gamma_{\rm b}$;
Col. [11] logarithm of the jet power in the form of radiation, [12] Poynting flux, [13]
bulk motion of electrons and [14] protons (assuming one cold proton
per emitting electron).
The spectral shape of the corona is assumed to be $\propto \nu^{-1} \exp(-h\nu/150~{\rm keV})$.
For all models we have assumed
a radius of the Broad Line region of $R_{\rm BLR} = 9.2\times 10^{17}$ cm,
a black hole mass of $2.8\times 10^9 M_\odot$ and
an accretion disk luminosity of $L_{\rm d}=9\times 10^{46}$ erg s$^{-1}$,
corresponding to $L_{\rm d}/L_{\rm Edd}=0.25$.
}
\label{para}
\end{table*}

\vskip 0cm

\section{Discussion}
\label{sec-discussion}

Figure~\ref{zoom} shows the X--ray data of B2~1023+25.
The X--ray SED data points were all 
absorption corrected and rebinned to have a $3\sigma$ detection in each bin.
Note that the {\it Swift}/XRT data are similar to those reported in S12 since variability
is negligible.   
Given the rapid variability seen routinely in blazars at many wavelengths, 
this lack of variability is evidence that the electrons
responsible for the hard X-ray emission have relatively small energies, 
and thus lose energy slowly. 
This is consistent with X--rays produced through the so--called External Compton process 
(Sikora et al.\ 1994), in which relatively cold electrons scatter broad line photons. 
Also the optical--UV emission is steady (see \S\ref{sec-grond}), for a completely 
different reason. 
This radiation is, in fact, emitted by the accretion disk, which is not expected to vary 
on short timescales. 

The radio part of the spectrum shows flux variability both at 5~GHz
and at high frequencies (see Fig.~2). A 43~GHz flux density
measurement was published recently by Frey et al. (2013), based on a
VLA A-configuration observation on UT 2002 June 19. A comparison to
their $f_{\nu}({\rm 43~GHz})=55\pm4$~mJy measurement to our CARMA
measurements clearly shows that the radio flux is variable in time, as
expected in blazars. The three radio data points obtained in this work
define a spectral index $\alpha_{\rm r}\sim$0.7 ($F(\nu)\propto \nu^{-\alpha_{\rm r}}$), 
steeper than $\alpha_{\rm r}\sim0.4$
reported by Frey et al. (2013). At least in part, this could be due to
the fact that observed frequency of 91~GHz corresponds to $\sim$570~GHz in
the source rest frame, likely sampling the optically thin part of the
synchrotron spectrum.

Thanks to the new X--ray flux results, we confirmed the extreme radio--loudness 
of B2~1023+25. 
During the sample selection (Sbarrato et al.\ 2013) we used the canonical 
radio--to--optical ratio to define its radio--loudness 
($R=F[5{\rm GHz}]/F[2500\rm\AA]\simeq 5200$), and  
this allowed to classify B2~1023+25 as the most radio--loud quasar of our sample. 
In addition to this, we now calculate the X--ray based radio--loudness 
$R_{\rm X}=\nu L_\nu[5{\rm GHz}]/L_{\rm X}[2-10~{\rm keV}]$,
using the X--ray fluxes and spectral indices listed in Table \ref{Xspec}. 
We obtain $\log R_{\rm X}=-0.65$ and $\log R_{\rm X}=-0.72$ 
(for fixed and free $N_{\rm H}$, respectively). 
Both values confirm the extreme radio--loudness of B2~1023+25 
according to the calibration introduced by Terashima \& Wilson (2003), 
which classifies quasars as radio-loud if they have $\log R_{\rm X}>-4.5$.

In S12, we derived a set of parameters that reproduced the observed SED
and suggested to classify B2~1023+25 as a blazar
(bulk Lorentz factor $\Gamma=14$, jet viewing angle $\theta_{\rm v}=3^\circ$). 
We fit the new observations using the model described in Ghisellini \& Tavecchio (2009).
Since this is a one--zone model, which assumes that the emitting region is 
rather compact, it cannot account for radio emission, which in the considered 
region is severely self--absorbed. 
In this model, both $\theta_{\rm v}$ and $\Gamma$ are free parameters, and
can be chosen independently.
Because of the hard and bright X--ray spectrum shown by B2~1023+25, 
we find a small value of $\theta_{\rm v}$ and a large Doppler boosting (i.e.\ large $\Gamma$). 
We find $\theta_{\rm v}<1/\Gamma$, as is typical of known blazars. 
Hence we confirm B2~1023+25 as a blazar (see \S\ref{bestfit}).

Because of the limited statistics in the X--ray spectrum, we investigate the range of
models consistent with the uncertainties.  As noted above, depending on how
the spectrum is modeled, the intrinsic continuum may be softer and overall fainter. 
This case implies a larger
value of $\theta_{\rm v}$ and a somewhat smaller value of $\Gamma$ (see \S\ref{misalign}).
The jet viewing angle, $\theta_{\rm v}$,  associated with this limiting solution is an
upper limit.  Since this model is also characterized by less Doppler boosting, 
it corresponds to a larger intrinsic luminosity relative to the SED corresponding
to the X-ray best fit parameters.
We consider then  ``re-orienting" the jet to a typical blazar viewing angle (i.e. $\sim3^\circ$) 
and we check if the corresponding  SED resembles the one of a typical powerful blazar
seen at lower redshift. We then use this to check the reliability of the obtained solution;
that is, we require that, if the jet were pointed toward us, the solution would show reasonably 
similar properties to the blazar sample. 

In our modeling we keep the parameters associated with the thermal emission from the accretion disk fixed. 
We assume a black hole mass $M_{\rm BH}=2.8\times10^9M_\odot$ 
and an accretion disk luminosity $L_{\rm d}=9\times10^{46}\ergs$, as 
derived in S12. 
Note that varying the black hole mass value inside the formal confidence range 
($M_{\rm BH}=1.8-4.5\times10^9M_\odot$)
does not change the results of our SED modeling.

\subsection{Best fit: small viewing angle, large bulk Lorentz factor}
\label{bestfit}

In our best fit model we find a set of parameters consistent with those from S12  
($\Gamma=13$, $\theta_{\rm v}=3^\circ$).
We report these in the first line of Table \ref{para} as the best fit to the broad band SED.
The case in Table \ref{para} corresponds to
the best fit to the X-ray data with $N_{\rm H}$ free to vary.   
Using $N_{\rm H}$  
fixed to the Galactic value yields a harder spectrum and therefore
an even more extreme blazar classification.
The model (blue solid line in Figures \ref{zoom}, \ref{sed1}  
and \ref{sed2}) describes a typical blazar, with the viewing angle 
smaller than the jet beaming angle ($\theta_{\rm v}<1/\Gamma$), firmly 
classifying B2~1023+25 as a blazar.

If $\theta_{\rm v}<1/\Gamma$ as in our best-fit model,  the number of sources similar to B2~1023+25 
but with the jet oriented outside of our line of sight
is $2\Gamma^2=338(\Gamma/13)^2$  
in the portion of the sky covered by SDSS+FIRST 
(Ghisellini et al.\ 2010a, Volonteri et al.\ 2011, Ghisellini et al.\ 2013).
Since the combined  SDSS and FIRST surveys (from which we selected B2~1023+25)
cover together 8770 deg$^2$, 
this implies that in the whole sky there must be at least $\sim1550$ 
sources that share the same intrinsic properties of B2~1023+25.
Since the co--moving volume in the redshift frame $5<z<6$ is $\sim380$ Gpc$^3$, 
we can conclude that there must be at least four radio--loud AGN similar to B2~1023+25 
per Gpc$^3$.  
Albeit extrapolating from  a sample of one, this would imply 
the presence  in the redshift bin $5<z<6$ 
of at least four supermassive black holes per Gpc$^3$, 
with a black hole mass of $M_{\rm BH}\sim10^9M_\odot$, 
hosted in jetted systems.

\subsection{Slightly misaligned jet}
\label{misalign}

\begin{figure*}
\vskip -1 true cm 
\epsscale{1.4}
\plotone{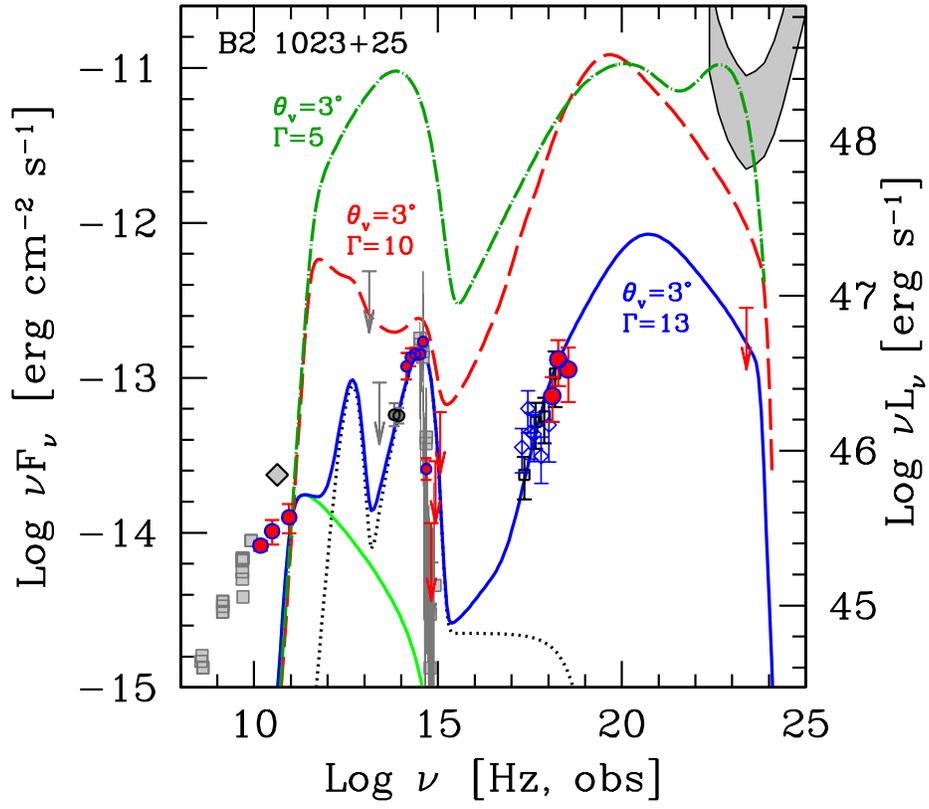}
\vskip -0.5 true cm
\caption{
SED of B2~1023+25 with data as in Figure \ref{sed1}.
The curved (grey) stripe corresponds to the sensitivity of {\it Fermi}/LAT 
after 1 yr of operation (5$\sigma$, lower bound) and of 3 months 
(10$\sigma$, upper bounds).
The solid (blue) line is the model with parameters as in the first row of Table \ref{para}. 
The dashed (red) line is the model with parameters as in the second row of Table \ref{para} 
(i.e. $\theta_{\rm v}=8^\circ$), but ``re--oriented" at $\theta_{\rm v}=3^\circ$, as labelled.
The dot--dashed (green) line is the model with parameters as 
in the third line of Table \ref{para} (i.e. $\theta_{\rm v}=20^\circ$), 
but ``re--oriented" at $\theta_{\rm v}=3^\circ$, as labelled.
} 
\label{sed2}
\end{figure*}

Figure \ref{zoom} shows that the X--ray data have large error bars. 
A softer X--ray spectrum cannot be excluded at 90\%  confidence (see Table \ref{Xspec}). 
A softer spectrum implies a larger viewing angle and therefore
a somewhat less extreme bulk Lorentz factor.
Specifically, a viewing angle of $\theta_{\rm v}=8^\circ$ with $\Gamma=10$
together with the other parameters in the second line of Table \ref{para}
still reproduce the broad--band data.
The model (red solid line in Figures \ref{zoom} and \ref{sed1}) 
represents an alternative interpretation consistent with the X--ray data points 
at 90\% level of confidence. 
This viewing angle, slightly larger than the jet beaming angle $1/\Gamma$,
still classifies B2~1023+25 as a blazar.
As a consistency check, we test 
how an object with the same instrinsic (comoving) properties would look if 
oriented at  $\theta_{\rm v}=3^\circ$, i.e.\ at $\theta_{\rm v}<1/\Gamma$.
The re--oriented model is shown in Figure \ref{sed2} by the dashed line 
(red dashed line in the electronic version). 
The resulting X-ray flux would be unusual but not unprecedented, 
being very similar, for example, to GB 1428+4217 ($z=4.72$, Worsley et al.\ 2004), 
although the latter shows a synchrotron hump much dimmer than 
our ``re--oriented" B2~1023+25.
We conclude that $\theta_{\rm v}=8^\circ$ is the largest possible viewing angle  
consistent with the {\it NuSTAR} data.

If $\theta_{\rm v}>1/\Gamma$, as in the above case, the number of sources similar to B2~1023+25 
but with their jet oriented in random directions is $1/(1-\cos\theta_{\rm v})$.
Hence, with $\theta_{\rm v}=8^\circ$, in the portion of sky covered by SDSS+FIRST, 
there would be 103 AGN analogous to B2~1023+25, and in the whole sky there 
would be 469  (i.e.\ $\sim$1.2 object per Gpc$^3$).
Even in this limiting case the number of extremely massive black holes in jetted systems 
in the redshift bin $5<z<6$ is cosmologically significant.

\subsection{Can the jet be at $20^\circ$ from our line of sight?}

Frey et al.\ (2013) claim that a viewing angle of at least $\sim20^\circ$ 
with a bulk Lorentz factor $\Gamma\sim15$ can be inferred  
for B2~1023+25 from published 5 GHz Very Long Baseline Interferometry (VLBI) imaging data. 
We therefore attempt to fit our X--ray data with $\theta_{\rm v}=20^\circ$ to test this hypothesis.
We find that the data are not consistent with both a large viewing angle and 
a large Lorentz factor, i.e.   
the values obtained by Frey et al.\ (2013).
In this case the corresponding Doppler factor 
$\delta=[\Gamma(1-\beta\cos\theta_{\rm v})]^{-1}\sim$1, and the intrinsic
jet power becomes huge, to account for the observed X--ray flux 
(i.e. $P_{\rm jet}\sim 10^{50}$ erg s$^{-1}$). 
In addition, the fit to the observed data is poor.
The maximum bulk Lorentz factor providing a good fit to the 
{\it NuSTAR} and broad band data has a viewing angle of
$\theta_{\rm v}=20^\circ$ is $\Gamma=5$ 
(along with the parameters in the third line of Table \ref{para}).
This model is shown in Figure \ref{sed1} as the dot--dashed line 
(green dot--dashed line in the electronic version).  
The corresponding beaming factor is $\delta\sim2.5$. 
Such a modest beaming factor implies that intrinsic luminosity would be very high. 
This would imply a class of objects with an extreme intrinsic luminosity. 
If such objects existed (at any redshift), we should see a few of them pointing at us. 
For illustration, we then ``re--orient" 
B2~1023+25 to $\theta_{\rm v}=3^\circ$ (see Figure \ref{sed2}).
Similar SEDs have never been observed, at any redshift.
All powerful blazars observed so far have the Compton component 
dominating the overall SED, contrary to what shown in Figure \ref{sed2}.  
We therefore believe that it is highly unlikely that B2~1023+25 can be described with 
$\theta_{\rm v}=20^\circ$ in the high--energy emitting region.
We cannot exclude the possibility that the jet bends between 
the X--ray and the radio emitting regions.
In this case, it is possible that the large scale jet (i.e.\ radio emission) is seen 
at a larger viewing angle than the compact jet. 

Furthermore, consider that the 5 GHz VLBI observations analyzed by Frey et al. (2013)
correspond to a rest frame frequency of 31.5 GHz.
At this frequency, all the VLBI components but the very inner core are emitting
thin synchrotron radiation. 
Since the brightness temperature of a synchrotron source peaks at the 
self--absorption frequency, we conclude that all the brightness temperature
of the resolved components are lower limits.
For the core, Frey et al. (2013) indeed performed a fit with a resolved plus
an unresolved component. 
It is very likely that the resolved core is optically thin (thus giving
a lower limit to the brightness temperature), while the unresolved core
gives a lower limit because of the upper limit on the size.
As a consequence, the derived Doppler factors are all lower limits, 
and the derived viewing angles are all upper limits.

\section{Conclusions}
\label{sec-concl}

We selected B2~1023+25  as the best $z>4$ blazar candidate from the SDSS+FIRST 
quasar catalog, and we classified it as a blazar as the result of a {\it Swift}/XRT 
ToO observation (S12).

Here we report simultaneous {\it NuSTAR} and {\it Swift} observations 
to improve the broad band X-ray spectrum and further cement the blazar classification.
We use the improved data to determine the jet orientation and 
the relativistic boosting factor.
Simultaneous GROND data are important to check for possible variability. From the comparison
with our first data, B2~1023+25 does not show variability 
in either its thermal or non--thermal emission.  

We fit the broadband SED with the model described in Ghisellini \& Tavecchio (2009), 
focusing on the X--ray energy band as an important constraint, and we analyze in detail 
the non--thermal jet emission of the source. 
We confirm that B2~1023+25 is an extremely radio--loud quasar, with a jet 
oriented very close to our line of sight, and hence the Doppler boost is large. 
Our SED modeling indicates a small viewing angle ($\theta_{\rm v}=3^\circ$)
associated with a large bulk Lorentz factor of $\Gamma=13$. 
To account for the large X--ray data uncertainties, we tested solutions 
with larger viewing angles.  
A model with $\theta_{\rm v}=8^\circ$ and $\Gamma=10$ cannot be excluded at 
the 90\% uncertainty level by the data.
A viewing angle larger than this is not consistent with the data, and the resulting solution 
provides a lower limit to the real X--ray spectrum. 
Therefore, 
B2~1023+25 shows a jet orientation and a Doppler boosting that lead us to firmly classify it 
as the second most distant blazar known ($z=5.3$). 
This implies the presence in the SDSS+FIRST survey of several  
detectable radio--loud sources with jets oriented in other directions. 
We however have not been able to identify such objects in the survey; there are only four other 
radio--detected quasars in the SDSS+FIRST 
sample at $z>5$.   Although the statistics are small and possibly not constraining, 
the apparent inconsistency is addressed in Volonteri et al.\ (2011).

B2~1023+25 is the first object at $z>5$ detected by {\it NuSTAR}, and 
confirms that {\it NuSTAR} is a very useful instrument to deepen our knowledge of the 
high--redshift X--ray universe. 
Specifically, {\it NuSTAR} could be an ideal tool to continue the $z>4$ 
blazar hunt.
With the study of this single object, indeed, we were able to estimate, albeit with
large uncertainty, how many extremely massive black holes in jetted sources 
are present at $5<z<6$.   
This constrains the mass function of heavy black holes in jetted systems 
as a function of redshift, which provides a complimentary constraint 
for surveys of radio--quiet AGN 
(see e.g.\ Ghisellini et al.\ 2013; Willott et al.\ 2010). 
The confirmation that B2~1023+25 is a blazar strengthens the suggestion of 
Ghisellini et al.\ (2013) that there are two epochs 
of heavy black holes formation: radio--loud objects preferentially 
form their $M_{\rm BH}>10^9M_\odot$ black holes at $z\sim4$, while radio--quiet 
quasars are formed at $z\sim2$.

\section*{Acknowledgements}
We would like to thank the anonymous referee for useful
comments.
We acknowledge financial support from the ASI-INAF grant I/037/12/0.
This work was supported under NASA Contract No. NNG08FD60C, and
made use of data from the {\it NuSTAR} mission, a project led by
the California Institute of Technology, managed by the Jet Propulsion
Laboratory, and funded by the National Aeronautics and Space
Administration. We thank the {\it NuSTAR} Operations, Software and
Calibration teams for support with the execution and analysis of
these observations.  
This research has made also use of the {\it NuSTAR} Data Analysis Software
(NuSTARDAS) jointly developed by the ASI Science Data Center (ASDC,
Italy) and the California Institute of Technology (Caltech, USA).
The scientific results reported in this article are based in part  on observations 
made by the {\it Chandra X-ray Observatory} and published previously in cited articles.
This publication makes use of data products from the {\it Wide--field
Infrared Survey Explorer}, which is a joint project of the University
of California, Los Angeles, and the Jet Propulsion
Laboratory/California Institute of Technology, funded by the National
Aeronautics and Space Administration.
Part of this work is based on archival data, software or on--line services 
provided by the ASI Data Center (ASDC).
This research has made use of the XRT Data Analysis Software (XRTDAS)
developed under the responsibility of the ASI Science Data Center (ASDC), Italy.
Part of the funding for GROND
(both hardware as well as personnel) was generously granted from the Leibniz
Prize to Prof. G. Hasinger (DFG grant HA 1850/28--1).
Support for CARMA construction was derived from the Gordon and Betty Moore Foundation, 
the Kenneth T. and Eileen L. Norris Foundation, the James S. McDonnell Foundation, 
the Associates of the California Institute of Technology, the University of Chicago, 
the states of California, Illinois, and Maryland, and the National Science Foundation. 
Ongoing CARMA development and operations are supported by the 
National Science Foundation under a cooperative agreement, 
and by the CARMA partner universities.
The operation of the OVRO 40--meter telescope is supported by 
NASA awards NNX08AW31G and NNX11AO43G, 
and NSF awards AST-0808050 and AST-1109911.
M.~B.\ acknowledges support from the International Fulbright Science and
Technology Award.


\label{lastpage}
\end{document}